\documentclass[twocolumn,resetfootnote]{aastex701}

\usepackage[utf8]{inputenc}
\usepackage{pifont}
\usepackage{amsmath}
\usepackage{stmaryrd}
\usepackage{amssymb}
\usepackage{wasysym}
\usepackage{amsbsy}
\usepackage{soul}
\usepackage{xcolor}

\usepackage{graphicx}
\usepackage{threeparttable}

\shorttitle{2015-2025 Optical Variability of OJ 287}
\shortauthors{Gupta et al.}

\begin{document}

\title{Multi-band optical photometric variability of the blazar OJ 287 from 2015 to 2025}

\correspondingauthor{Alok C. Gupta, Karan Dogra, Lang Cui}
\email{acgupta30@gmail.com, karandogra987@gmail.com, cuilang@xao.ac.cn} 

\author[0000-0002-9331-4388]{Alok C.\ Gupta}
\affiliation{Xinjiang Astronomical Observatory, Chinese Academy of Sciences, 150 Science 1-Street, Urumqi 830011, China}
\affiliation{Aryabhatta Research Institute of Observational Sciences (ARIES), Manora Peak, Nainital 263001, India}
\affiliation{Abastumani Observatory, Mt. Kanobili, 0301 Abastumani, Georgia}
\email{acgupta30@gmail.com}

\author[0009-0007-3214-602X]{Karan Dogra}
\affiliation{Aryabhatta Research Institute of Observational Sciences (ARIES), Manora Peak, Nainital 263001, India}
\email{karandogra987@gmail.com}

\author{Mark Kidger}
\affiliation{European Space Astronomy Centre, European Space Agency, E-28691 Villanueva de la Ca$\tilde{n}$ada, Madrid, Spain}
\affiliation{Variable Star Section, British Astronomical Association, Villanueva del Pardillo, 28229 Madrid, Spain}
\affiliation{Observadores de Supernovas (ObSN), Villanueva del Pardillo, 28229 Madrid, Spain}
\affiliation{The Astronomer, Villanueva del Pardillo, 28229 Madrid, Spain}
\email{mark.kidger@ou.ac.uk}
 
\author[0000-0001-8580-8874]{Mauri J. Valtonen}
\affiliation{FINCA, University of Turku, Turku, Finland}
\affiliation{Department of Physics and Astronomy, University of Turku, Turku, Finland} 
\email{mvaltonen2001@yahoo.com}

\author[0000-0002-1029-3746]{Paul J. Wiita}
\email{wiitap@tcnj.edu}
\affiliation{Department of Physics, The College of New Jersey, 2000 Pennington Rd., Ewing, NJ 08628-0718, USA}
 
\author[0000-0001-6890-2236]{Pankaj Kushwaha}
\email{pankaj.kushwaha@iisermohali.ac.in}
\affiliation{Department of Physical Sciences, IISER Mohali, Knowledge City, Sector 81, Manauli PO SAS Nagar, Punjab 140306 India}

\author[0000-0003-4147-3851]{Sergey S.\ Savchenko}
\email{savchenko.s.s@gmail.com}
\affiliation{St. Petersburg State University, 7/9, Universitetskaya nab., 199034 St. Petersburg, Russia}
 
\author[0000-0002-0319-5873]{Sofia O.\ Kurtanidze}
\affiliation{Abastumani Observatory, Mt. Kanobili, 0301 Abastumani, Georgia}
\affiliation{Landessternwarte, Zentrum f$\ddot{u}$r Astronomie der Universit$\ddot{a}$t Heidelberg, K$\ddot{o}$nigstuhl 12, 69117 Heidelberg, Germany}
\email{sofiakurtanidze@gmail.com}
 
\author[0000-0001-6158-1708]{Svetlana G. Jorstad}
\email{jorstad@bu.edu}
\affiliation{Institute for Astrophysical Research, Boston University, 725 Commonwealth Avenue, Boston, MA 02215, USA}
\affiliation{St. Petersburg State University, 7/9, Universitetskaya nab., 199034 St. Petersburg, Russia}
 
\author[0000-0001-7396-3332]{Alan P. Marscher}
\email{marscher@bu.edu}
\affiliation{Institute for Astrophysical Research, Boston University, 725 Commonwealth Avenue, Boston, MA 02215, USA}

% ...

\begin{abstract}
\noindent
We present the most densely sampled multi-band optical photometric observations of the peculiar BL Lacertae object OJ 287 from 2015 to 2025 with a focus on its optical activity on diverse timescales. We present a total of 2296, 10927, 11484, and 2982 data points in B, V, R, and I bands, respectively.  The densely sampled observations allow us to keep track of the source evolution that it has exhibited since the start of the predicted major optical flaring activity at the end of 2015. The study reveals clear and persistent bluer when brighter trends in both the long-term and short-term variations. Different bands were cross-correlated with discrete correlation functions, which peak at zero lag, implying co-spatial emission. Using  eight optical spectra in the low flux states of OJ 287 taken from 2017 October 21 to 2017 November 22, from Steward Observatory, we estimate the central black hole mass to be at least 3.89 $\times \ \rm{10}^{9} \ \rm{M}_{\odot}$ from the [O III] line width. The emission mechanism of the binary black hole blazar, and its possible implication in various aspects of multi-messenger astronomy are briefly discussed. 
\end{abstract}


\section{Introduction}
\label{sec:intro}
\noindent
Blazars are jetted active galactic nuclei (AGNs), that are traditionally designated as being radio-loud, with a large-scale, powerful, relativistic jet of plasma directed nearly toward our line of sight. They are characterized by a strong and rapidly variable continuum that spans the entire accessible electromagnetic (EM) spectrum from radio to $\gamma$-rays
\citep[e.g.][]{2021Univ....7..421D,2017ICRC...35..650B} with  high and variable polarized radio and optical continua; some exhibit recently detected variable X-ray polarization too \citep[e.g.][]{2023ApJ...957L..11G,2023PASJ...75....1I,2023NatAs...7.1245D}. 
The broadband multi-wavelength (MW) continuum revealed a new characteristic feature of blazar emission -- a broad bi-modal spectral energy distribution \citep[SED;][]{1998MNRAS.299..433F,2010ApJ...716...30A}, primarily of non-thermal nature. Studies of these two humps are critical in investigations of the emission mechanisms and the underlying physical conditions \citep[e.g.][]{2019NatAs...3...88G,2018MNRAS.479.1672K,2018MNRAS.473.1145K,2022Galax..10...35P,2022MNRAS.509.2696S,2025Univ...11...84K,2025MNRAS.544.2455N,2025MNRAS.540..582H}. Given the ubiquity of variations in all bands, these explorations are greatly facilitated by simultaneous broadband monitoring. 
The low-energy part of the SED emission, extending from radio to up to X-rays, reaching a maximum between the infrared (IR) and X-ray bands is widely accepted to be the synchrotron emission from ultra-relativistic non-thermal electrons in the relativistic jet. The origin of the high-energy part of the SED is still unclear, as both leptonic and hadronic models reproduce it reasonably well within the available observational constraints. The leptonic emission scenarios attribute it to inverse Compton (IC) scattering off the synchrotron emitting electrons of a combination of soft photon fields arising both within and external to the jet \citep[e.g.][]{,2022MNRAS.509.2696S}. The alternative hadronic scenarios reproduce this high energy emission via proton synchrotron and/or cascades resulting from the interactions of relativistic protons with proton and photon fields \citep[e.g.][]{2019NatAs...3...88G}.\\

\begin{table*}
\caption{The observatories, data collection periods, and other relevant details.}
\centering
\begin{tabular}{lcclccccc}
\hline
\hline
Observatory/ & MJD  & MJD   & Origin  & Aperture & Filters & Data points &  Marker & Color \\
Observer& Start & Stop &&(cm)&&$(N_{obs})$&&(in Fig.\ 1) \\
\hline
\hline
Abastumani& 58126.43& 60430.26& Georgia& 70& R& 707                           &$\pentagon$ &Slategrey  \\ 
A. Darriba$^d$& 58103.55& 59209.66&Spain& 35.6& V& 76                                     &$\square$ &Blue  \\ 
ARIES$^a$& 58562.15& 59197.48& India& 104/130&BVRI &11,11,11,11               &$+$ &Lime  \\  
ASV$^b$& 58144.37& 59700.35& Serbia& 60/140&BVRI &86,49,83,85                 &$\Diamond$ &Green  \\   
Belogradchik& 58228.26& 60414.33& Bulgaria& 60& BVRI&55,60,60,53              &$\pentagon$ &Gold  \\   
Crimean& 57524.27& 59863.59& Russia& 70& BVRI&144,158,464,204                 &\ding{54} &Teal  \\ 
Mark Kidger$^*$& 57356.42& 61139.82& International& Various& BVRI&1104,9024,5973,1408   &$\lozenge$ &Purple  \\ 
Osaka& 57295.29& 60114.97& Japan&150 &B &294                                     &$\hexagon$ &Tan  \\ 
Perkins& 60228.53& 61030.44& United States& 180& BVRI& 285,294,166,282& $\bigcirc$& Crimson \\
Rozhen& 57369.59& 58993.30& Bulgaria& 50/70/200&BVRI &53,62,85,63             &$\triangledown$ &Black  \\   
SATU$^{f}$& 57816.00 &59558.00 & West-Indies& 40& R& 15                         &$\octagon$ &Salmon  \\
SMARTS$^c$& 57353.84& 57857.55& Chile& 130/150& BVR&111,110,111               &$\triangle$ &Dark-violet  \\  
St. Petersburg& 57513.32& 59938.62& Russia& 40&BVRI &66,80,314,202            &$\triangleleft$ &Olive  \\   
Steward$^e$& 58130.85& 58292.65& United States& 154/230& VR& 34,34            &$\bigcirc$ &Cyan  \\   
Weihai& 58121.27& 58994.03& China&100 & VRI& 115,121,114                          &$\bigcirc$ &Magenta  \\   
\hline
Paper 1$^g$& 57286.01& 57529.48& International& Various& BVRI&40,190,437,45       &\ding{58} &Brown  \\ 
Paper 2$^h$& 57309.32& 58103.82& International& Various& BVRI&46,664,840,515      &$\star$ &Orange  \\ 
Paper 3$^i$& 57284.00& 59993.28& International& Various& R& 1122                  &$\triangleright$ &Pink  \\ 
\hline 
\label{tab:1}
\end{tabular} \\
\textbf{Notes.}\\
$^a$\emph{Aryabhatta Research Institute of Observational Sciences}\\
$^b$\emph{Astronomical station Vidojevica}\\
$^c$\emph{Small and Medium Aperture Research Telescope System}\\
$^d$\emph{Observatorio Astronomico Las Casqueras} \\
$^e$\emph{Steward Observatory Support of the Fermi Mission with 154 cm Kuiper and 230 cm Bok telescopes \citep{2009arXiv0912.3621S}}\\
$^{f}$\emph{St. Augustine - Tuorla Observatory, Trinidad and Tobago} \\
$^g$\emph{\cite{2017MNRAS.465.4423G}}, $^h$\emph{\cite{2019AJ....157...95G}}, $^i$\emph{\cite{2023ApJ...957L..11G}}\\
$^*$\emph{Multiple telescopes}\\

\end{table*}
 
\noindent
Blazars are taken to be the combination of flat-spectrum radio quasars (FSRQs) and  BL Lacertae objects (BL Lacs); both exhibit similar continuum behavior, with the primary difference being the optical spectrum. The former shows prominent broad emission line features, while in 
the latter, they are weak or these features are completely lacking. OJ 287 is a BL Lac type blazar at a redshift of $z = 0.304 $ \citep{1985PASP...97.1158S, 2010A&A...516A..60N}. It is one of the best-monitored blazars with an extensive pre-discovery light curve (LC), obtained from archival plates and early post-identification observations driven by its enigmatic dynamic variability at optical and radio bands \citep[][and references therein]{1985PASP...97.1158S}. Subsequently, the identification of recurring optical outbursts at $\sim12$-year intervals led to the popular interpretation of the source as a binary supermassive black hole (SMBH) system, with recurring outbursts attributed to the impact of the secondary on the accretion disk of the primary \citep[and references therein]{1988ApJ...325..628S,2025ApJ...992..110V}. The onset of the next set of flares, in 1994 and 1995, around the model-predicted schedule gave this hypothesis further credence and has driven substantial coordinated MW monitoring of OJ 287 \citep{1996A&A...305L..17S,1996A&A...315L..13S}. In its nearly regular cycle of outbursts, \citet{2006ApJ...646...36V} predicted that OJ 287 was due for another outburst season in 2006--2010. In an extensive optical photopolarimetric study of the source from 2005 to 2009, the predicted double-peaked outbursts were detected in December 2005 and December 2007, and it was suggested that both the double-peaked bursts and the evolution of the optical polarization position angle could be explained as a sign of resonant accretion of magnetic field lines, a ‘magnetic breathing’ of the disk \citep{2010MNRAS.402.2087V}. The next predicted double-peaked outbursts were detected in December 2015 \citep{2016ApJ...819L..37V} and July 2019 \citep{2020ApJ...894L...1L}. The July 2019 flare was detected using the Spitzer telescope, as the source was not available from ground-based optical/IR observations. OJ 287's MW behavior has been extensively observed since 2015 by various groups and the MOMO (Multiwavelength Observations and Modeling of OJ 287) collaboration \citep[e.g.][and references therein]{2016ApJ...819L..37V,2016ApJ...832...47B,2018MNRAS.473.1145K,2018MNRAS.479.1672K,2018ApJ...863..175G,2020ApJ...890...47P,2020ApJ...894L...1L,2020MNRAS.498L..35K,2021ApJ...923...51K,2021MNRAS.504.5575K,2021Univ....7..261K,2023ApJ...944..177K,2021A&A...654A..38P,2022MNRAS.509.2696S,2023ApJ...951..106B,2023ApJ...957L..11G}.
The other prominent interpretation of the recurrent flares attributes it to jet precession  \citep[e.g.][and references therein] {2018MNRAS.478.3199B} and it still continues to be invoked  \citep[e.g.][]{2020Univ....6..191B}, though these models primarily focus on explanation rather than prediction \citep[e.g. see][for a succinct summary of proposed models]{2020Galax...8...15K}. \\

\begin{figure*}
    \centering
    \includegraphics[scale=0.9]{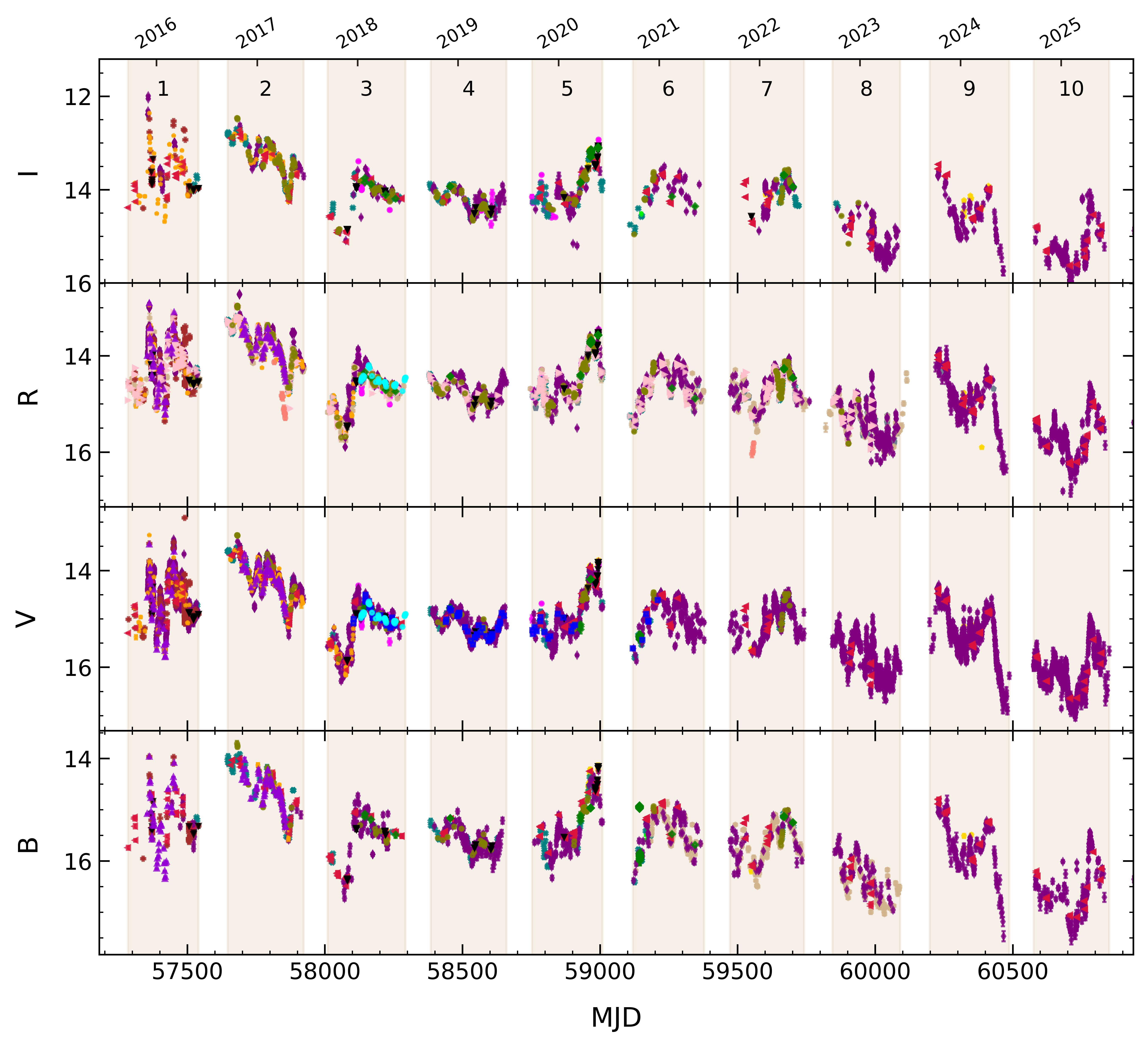}
    \caption{Multi-band optical light curve of OJ 287 from 2015 to 2025. The x-axis and y-axis represent modified Julian date (MJD) and optical magnitudes, respectively. All four light curves are divided into 10 segments due to seasonal gaps. The observatories providing the data are indicated by the shapes and colors of the points as defined in \autoref{tab:1}.}
    \label{fig:optLc}
\end{figure*}

\noindent
Despite the impact scenario binary-SMBH model being favored by current observational behavior, several aspects of the observations seem to be inconsistent with its predictions or claims. For example, the observed optical spectrum during flaring \citep[e.g.][]{2017MNRAS.465.4423G,
2018MNRAS.473.1145K,2009PASJ...61.1011S} does not show a flat (\(F_\nu \sim \nu^{0-0.2}\)) spectrum, which is characteristic of the thermal bremsstrahlung emission that is claimed for the outbursts. Furthermore, neither a MW activity with a simultaneous flare at X-rays and $\gamma$-rays \citep{2018MNRAS.473.1145K}, nor a systematic large swing of optical polarization \citep{2023ApJ...957L..11G} is expected during the flare in the case of the claimed thermal emission. Both of the latter behaviors indicate non-thermal broadband emission components. Additionally, the observed degree of optical polarization is still uncertain \citep[e.g.][]{2009PASJ...61.1011S}. Detection of both the expected impact-induced jet emission as well as the tidal-accretion has been claimed \citep[e.g.][]{2018MNRAS.475L..15G,2021ApJ...920...12H} and is also supported by previously available observations \citep[and references therein]{2001PASJ...53...79I}. However, the spectral nature is different \citep{2017ICRC...35..650B,2018MNRAS.479.1672K,2021ApJ...921...18K} to the jet emission considered to be normal for the source \citep[e.g.][]{2013MNRAS.433.2380K}. Since most of the characteristic properties of the models are associated with the optical bands, multi-band optical monitoring holds the key to addressing these issues, along with concurrent MW studies. \\

\begin{table*}
\caption{Total and Segment-wise variability amplitudes of OJ 287 from 2015-2025}
\begin{tabular}{ccccccccc}
\hline
\hline
\vspace{0.1cm}
& Band & $V_{amp}$  &   & Band & $V_{amp}$  &   & Band & $V_{amp}$\\
&  &  &  &  &  &  &  &  \\
\hline
\hline

&  B&  2.38 $\pm$ 0.01&  &  B& 1.88 $\pm$ 0.06 &  &  B&  2.00 $\pm$ 0.06\\
Segment 1&  V&  2.87 $\pm$ 0.02&  Segment 2&  V&  2.08 $\pm$ 0.04&  Segment 3&  V&1.97 $\pm$ 0.04  \\
&  R&  2.45 $\pm$ 0.02&  &  R&  2.57 $\pm$ 0.01&  &  R&2.03 $\pm$ 0.03  \\
&  I&  2.69 $\pm$ 0.02&  &  I& 1.79 $\pm$ 0.01&  &  I& 1.72 $\pm$ 0.02 \\

\hline

&  B& 1.11 $\pm$ 0.08&  &  B&  2.18 $\pm$ 0.06&  &  B& 1.56 $\pm$ 0.05\\
Segment 4&  V&  0.91 $\pm$ 0.05&  Segment 5&  V&1.98 $\pm$ 0.04  &  Segment 6&  V&1.42 $\pm$ 0.04  \\
&  R&  0.96 $\pm$ 0.03&  &  R&  2.05 $\pm$ 0.02&  &  R&1.58 $\pm$ 0.02  \\
&  I& 0.86 $\pm$ 0.07 &  &  I& 2.27 $\pm$ 0.03 &  &  I& 1.45 $\pm$ 0.03  \\

\hline

&  B& 1.49 $\pm$ 0.02&  &  B&  1.48 $\pm$ 0.08&  &  B& 2.67 $\pm$ 0.10\\
Segment 7&  V& 1.30 $\pm$ 0.08 &  Segment 8&  V&1.75 $\pm$ 0.05 &  Segment 9&  V&2.60 $\pm$ 0.07  \\
&  R&  1.98 $\pm$ 0.03&  &  R&1.84 $\pm$ 0.05 &  &  R&2.51 $\pm$ 0.11  \\
&  I&  1.31 $\pm$ 0.02&  &  I& 1.41 $\pm$ 0.03 &  &  I& 2.28 $\pm$ 0.09  \\

\hline

&  B& 2.10 $\pm$ 0.09 &&B&3.83$\pm$0.09&&&\\
Segment 10&  V& 2.09 $\pm$ 0.07&Total&V&4.13$\pm$0.06&&& \\
&  R&  2.25 $\pm$ 0.08&Light curve&R&4.14$\pm$0.07&&& \\
&  I&  2.14 $\pm$ 0.08 &&I&4.22$\pm$0.06&&& \\

\hline
\label{tab:2}
\end{tabular}
\end{table*}

\noindent
OJ~287 is monitored most thoroughly in optical bands, particularly V and R -- the legacy EM window and the optical window are where the appearance of recurring outbursts is reported. Apart from the timing information, the polarization and spectral shape from multi-band optical monitoring can provide the most crucial test of the model's spectrum, encoded in the color-flux variations \citep[e.g.][]{2024A&A...688L..16G,2020Galax...8...15K}. Though the binary-impact driven thermal flare is limited most likely to and around the expected recurrent flare, the rest of the activity is driven entirely by the jet -- characterized by a non-thermal power-law spectrum, imprinted in the color variation which is a direct tracer of this and any additional spectral component \citep[e.g.][]{,2018MNRAS.479.1672K}.
The color variability of blazars is often divided into two subcategories: bluer when brighter BWB and redder when brighter (RWB). The former is generally associated with the BL Lac objects \citep{2006A&A...450...39G, 2013MNRAS.432.1189Z, 2015A&A...573A..69W}, and the latter with the FSRQs \citep{2008A&A...491..755R, 2012ApJ...756...13B}. A recent study conducted by \cite{2025ApJS..276....1D} revealed that the FSRQ 3C 454.3 follows an RWB trend up to a particular brightness value, after which the color behavior can be RWB, BWB, or achromatic. Nonetheless, some cases have been reported where the trends were interchanged \citep{2011A&A...528A..95G} or the objects under study display achromatic behavior \citep{2023MNRAS.522..102R} instead. Most of the previous studies were done either on short timescales or did not have a sufficient number of observations to confidently constrain this behavior, which motivated us to study the long-term as well as short-term color behavior of OJ~287. \\
\\
In this work, we present the most extensive and up-to-date study of the multi-band optical flux and spectral variability of the source from September 2015 to April 2025, the expected time of the onset of the first portions of an impact-model predicted recurrent optical outburst \citep{2016ApJ...819L..37V}. This is a continuation of our ongoing project on OJ 287 aimed at presenting a dense optical follow-up between the \(\sim 12\)-yr optical outbursts in order to investigate and understand the behavior of this source and compare the emission with traits considered typical of the source and then allow us to make comparisons with proposed models \citep{2017MNRAS.465.4423G, 2019AJ....157...95G, 2022ApJS..260...39G, 2023ApJ...957L..11G,    2018MNRAS.479.1672K, 2018MNRAS.473.1145K, 2021ApJ...921...18K, 2022JApA...43...79K, 2024ApJ...960...11K}. \\
\\ 
The present work is divided into six sections, with \autoref{sec:data} briefly describing the optical observations and data reduction. In  \autoref{sec:analysis} we give the results of our multi-band optical data analysis, with \autoref{sec:BH_mass} explaining the BH mass calculation, followed by a discussion in \autoref{sec:discussion}. We close with a summary of the findings in \autoref{sec:summary}.

\section{Multi-band Optical Data} \label{sec:data}
\noindent
The multi-band optical data collected under these campaigns is shown in \autoref{fig:optLc}. The participating observatories, data collection periods, and other important information related to them are provided in \autoref{tab:1}. The image frames were first cleaned (e.g., using IRAF\footnote{IRAF is distributed by the National Optical Astronomy Observatories, which are operated by the Association of Universities for Research in Astronomy, Inc., under a cooperative agreement with the National Science Foundation.}), which involved bias correction, flat correction, and cosmic-ray removal, after which standard image reduction steps (e.g., using DAOPHOT II\footnote{Dominion Astronomical Observatory Photometry.}, ESO-MIDAS\footnote{European Southern Observatory - Munich Image Data Analysis System}, and MAXIM-DL6\footnote{CCD imaging and data reduction software by \emph{Diffraction Limited}},
etc.) were performed.\\
\\

Once we have performed cleaning and reduction of the raw science frames, differential photometry was performed to obtain BVRI magnitudes of OJ 287 using local standard stars 9, 12, and 13 in the blazar field \citep{2001AJ....122.2055G}. Most observatories used the Johnson–Cousins broadband BVRI filters for photometric observations; however, some amateur observations were obtained using unfiltered (CV, i.e., clear/white light) measurements. These CV data were subsequently transformed to the equivalent Johnson-Cousins V band. All of the data were corrected for Galactic extinction, with respective extinction values taken from NASA's extragalactic database.
As data from multiple telescopes were available, we performed inter-calibration between the different datasets. The offset between the light curves is calculated by identifying overlapping data points and then applying a mean shift, which sometimes is present  \citep{2017MNRAS.466.3762R, 2024A&A...682A.114M}. We found that the great majority of the offsets were well within the measurement uncertainties.

\begin{table*}
\centering
\caption{Fit parameters for color-magnitude diagrams}
\begin{tabular}{cccccccccc}
\hline
&&B-R&& &&&V-R&&\\
\hline
&Slope&Intercept&$r$&$\rho$&&Slope&Intercept&$r$&$\rho$\\
\hline
Segment 1& 0.0453 $\pm$ 0.0012& 0.3677 $\pm$ 0.0171& 0.312&0.405&    &      0.0187 $\pm$ 0.0108& 0.1878 $\pm$ 0.1557&  0.105& 0.092  \\
Segment 2& 0.0611 $\pm$ 0.0047& -0.1197 $\pm$ 0.0686& 0.396&0.419&   &      0.0516 $\pm$ 0.0102& -0.3706 $\pm$ 0.1436& 0.256& 0.296     \\
Segment 3& 0.1124 $\pm$ 0.0096& -0.8861 $\pm$ 0.1479 & 0.503&0.555&  &      0.0216 $\pm$ 0.0095& 0.0862 $\pm$ 0.1430& 0.174& 0.199     \\
Segment 4& 0.1686 $\pm$ 0.0081& -1.8098 $\pm$ 0.1262& 0.339& 0.317&  &      0.0944 $\pm$ 0.0127& -1.0468 $\pm$ 0.1926& 0.175& 0.223     \\
Segment 5& 0.1102 $\pm$ 0.0040& -0.8682 $\pm$ 0.0597& 0.710& 0.717&  &      0.0376 $\pm$ 0.0055& -0.1747 $\pm$ 0.0813& 0.398& 0.414     \\
Segment 6& 0.0679 $\pm$ 0.0060& -0.2077 $\pm$ 0.0915&0.337 & 0.387&  &      -0.0103 $\pm$ 0.0130& 0.5564 $\pm$ 0.1944& 0.053& 0.034     \\
Segment 7& 0.0739 $\pm$ 0.0046& -0.2846 $\pm$ 0.0718 & 0.408& 0.509& &      0.0422 $\pm$ 0.0124& -0.2164 $\pm$ 0.1852& 0.280& 0.305     \\
Segment 8& 0.0945 $\pm$ 0.0091& -0.5427 $\pm$ 0.1472& 0.330& 0.363&  &      0.0376 $\pm$ 0.0177& -0.1366 $\pm$ 0.2796& 0.359& 0.414     \\
Segment 9& 0.0304 $\pm$ 0.0063& 0.3407 $\pm$ 0.0972& 0.254& 0.258&   &      0.0209 $\pm$ 0.0087& 0.0743 $\pm$ 0.1309&  0.046& 0.021     \\
Segment 10& 0.1076 $\pm$ 0.0050& -0.9001 $\pm$ 0.0823& 0.570& 0.594& &      0.0573 $\pm$ 0.0074& -0.5095 $\pm$ 0.1168& 0.308&  0.381    \\
\hline
Total LC& 0.0720 $\pm$ 0.0042& -0.2500 $\pm$ 0.0502& 0.479& 0.495&&0.0311 $\pm$ 0.0004&-0.0565 $\pm$ 0.0055&0.232&0.234 \\
\hline
\label{CM_fit}
\end{tabular}
\end{table*}

\section{Variability Analysis}\label{sec:analysis}
\noindent
\autoref{fig:optLc} shows the optical flux behavior of OJ~287 between 2015 and 2025. Large amplitude variations can be seen in all four bands on similar timescales. As shown in \autoref{fig:optLc}, the light curve across all four bands is divided into 10 segments (or observing seasons) due to yearly seasonal gaps with other gaps for individual observatories caused by weather, maintenance, etc. All segments, as well as the full light curve, have been used to better study the source's variability. As blazars show variability in the entire EM spectrum and over diverse timescales, we need to quantify these features as they can help to elucidate the emission mechanism along with the physical properties of the emission region. The variability amplitude is computed for all light curve segments, the color-magnitude relationship is studied in detail, and the Z-transformed discrete correlation function is computed for all light curve pairs in \autoref{amp_var}, \autoref{color_var}, and \autoref{zdcf}, respectively.

\subsection{Variability amplitude}\label{amp_var}
\noindent
One of the simplest ways to measure variation is the variability amplitude. It is related to the minimum ($ m_{min}$) and maximum ($m_{max}$) reported magnitude within a given duration and the mean uncertainty of measurement ($\sigma$) as
 
\begin{equation}
    V_{amp} =  \sqrt{(m_{max} - m_{min})^2 - 2\sigma^2} .
\end{equation}
 
The error associated with variability amplitude is given by
\begin{equation}
    \Delta V_{amp} = {\sqrt{\frac{(m_{max}-m_{min})^2(\sigma^2_{max}+\sigma^2_{min})+4\sigma^2\Delta\sigma^2}{V_{amp}^2}}},
\end{equation}
where $\sigma_{max},\ \sigma_{min}$ and $\Delta\sigma$ are the uncertainties in $m_{max},\ m_{min}$ and $\sigma$. In order to have a proper track of how light curve variability varied over time, $V_{amp}$ has been calculated for all the light curve segments and bands, with the results given in \autoref{tab:2}.  

\begin{figure*}
\centering
\setlength{\tabcolsep}{2pt}  
\renewcommand{\arraystretch}{0.95}  
\begin{tabular}{cc}
\includegraphics[width=0.50\textwidth]{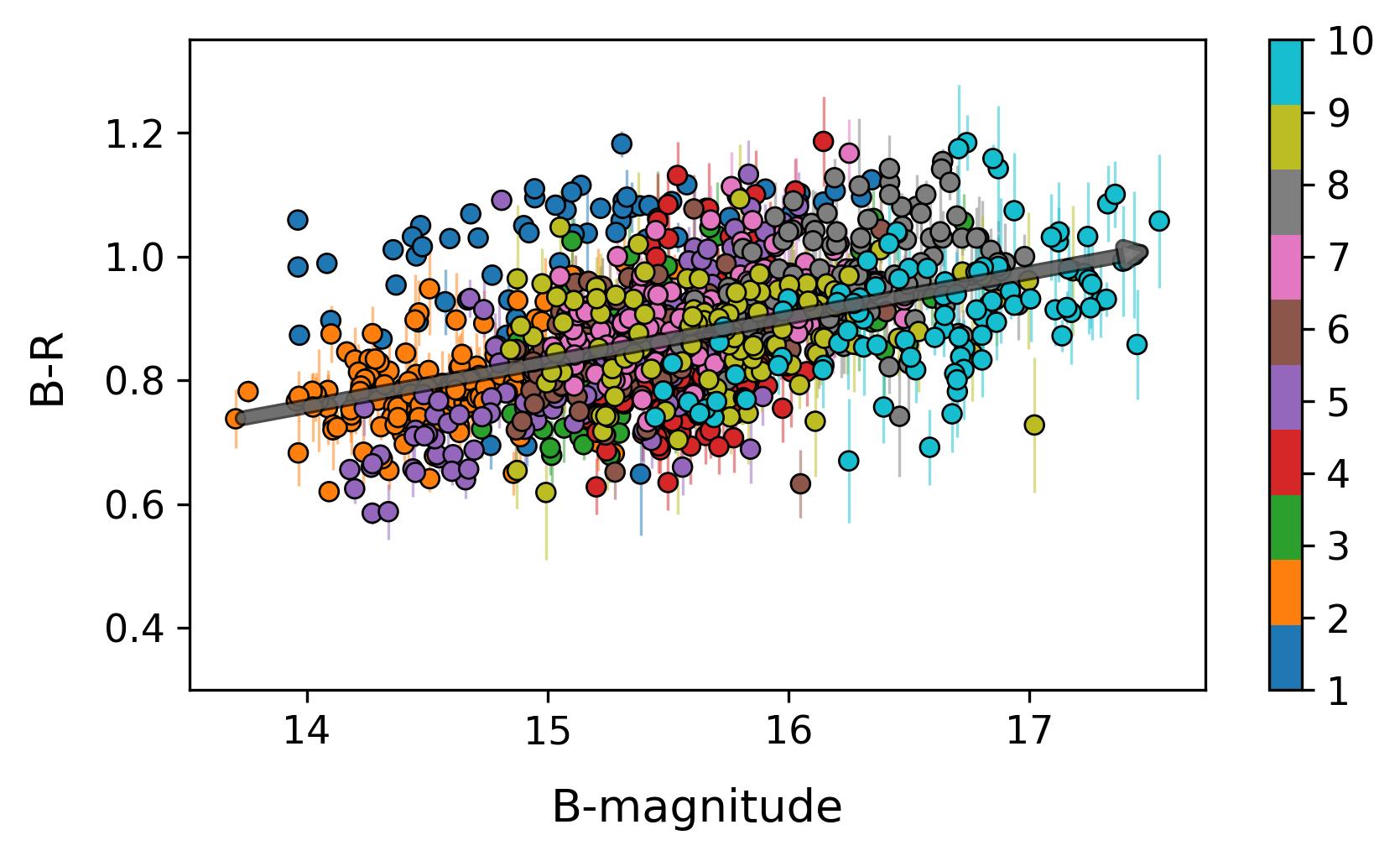} 
\includegraphics[width=0.50\textwidth]{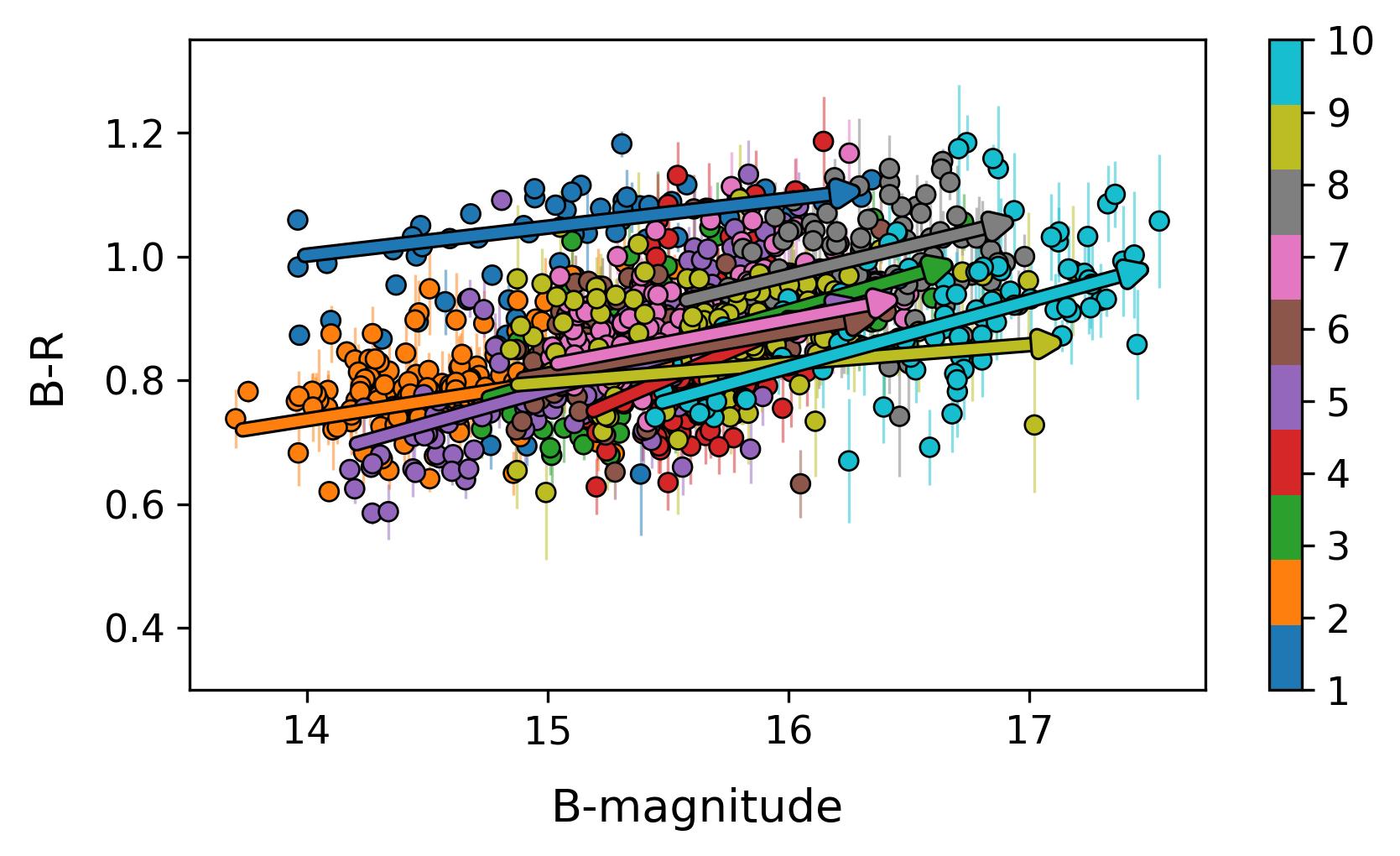}\\
\includegraphics[width=0.50\textwidth]{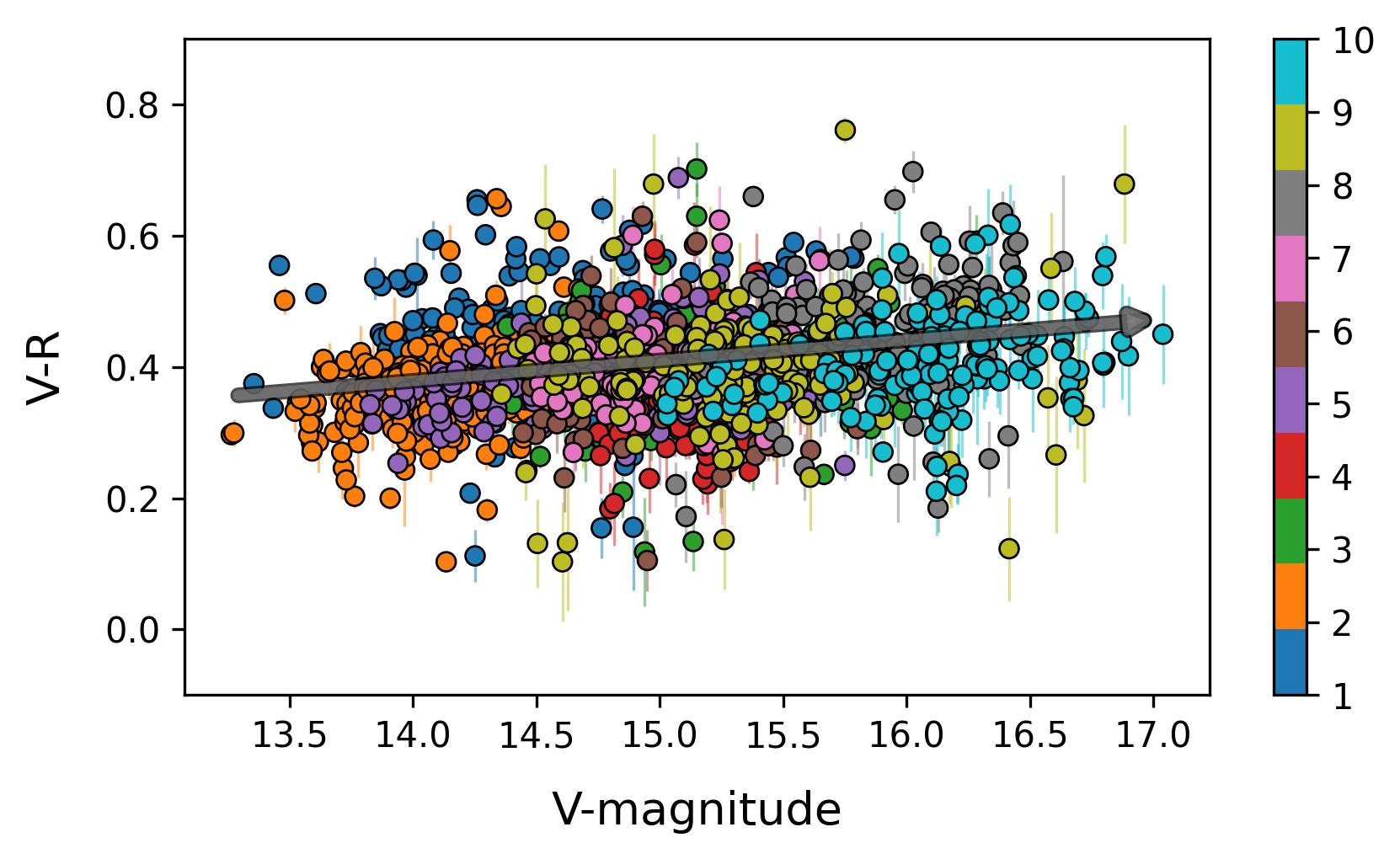} 
\includegraphics[width=0.50\textwidth]{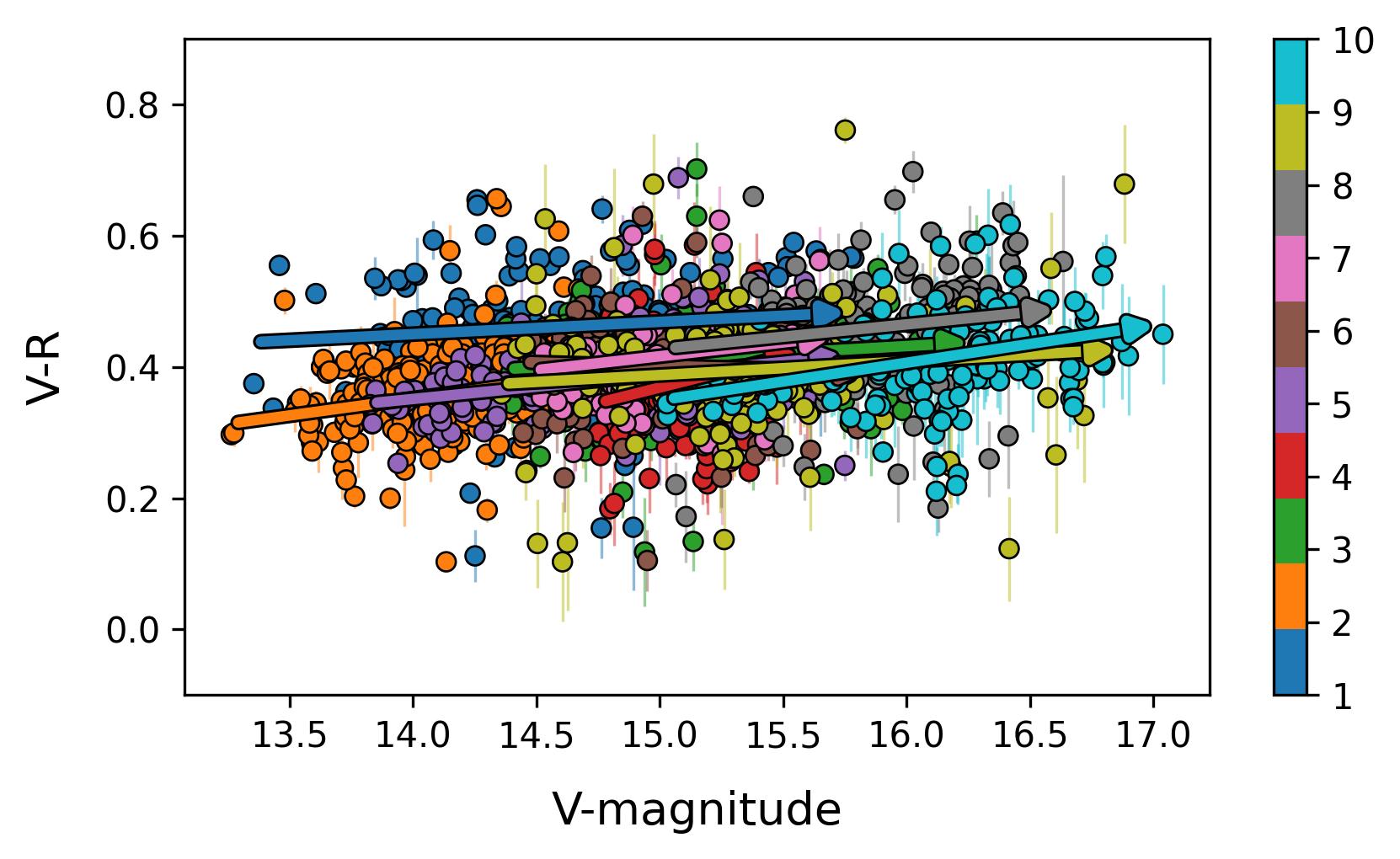}
\end{tabular}
\caption{The left panels in the figure show the color-magnitude diagrams with a linear fit to all data points, whereas the right panels show the fit to all ten segments.}
\label{fig:color}
\end{figure*}

\subsection{Flux and Color Variability}\label{color_var}
\noindent
Closely related to flux variability is color variability, often called the color-magnitude (CM) relation, which can hint at possible physical processes mainly in the jet of the blazar. To study the color variability, we plotted the B$-$R and V$-$R colors against the B and R magnitudes, respectively, as CM diagrams in \autoref{fig:color}. We chose B$-$R and V$-$R to plot differences between widely separated and close bands, respectively, as they have the largest number of color indices. 
To clearly track each segment's contribution in the CM diagrams, the segments are color-coded. Although a blazar's spectral energy distribution is dominated by non-thermal emission, imprints of quasi-thermal emission can also be noted, at least in the quiescent/low-state of the object under study.\\
\\
Apparently, an overall linear trend is evident in \autoref{fig:color} throughout the observation campaign and across all segments. To quantify this behavior, we performed linear fits, with their respective slopes, intercepts, Pearson coefficients, $r$, and Spearman  correlation coefficients, $\rho$, given in \autoref{CM_fit}. We obtained 1425 B$-$R color indices and 2110 V$-$R color indices, with averages of 0.934 and 0.393, respectively. As can be seen in \autoref{fig:color}, the BWB trend becomes more evident as the difference in color between bands increases. Apart from color-magnitude variations, color versus MJD variations have also been studied, as shown in \autoref{fig:color_vs_JD}, which reveals a slow variation over the observing seasons fitted using a cubic polynomial of the form $ax^3 +bx^2+cx+d$. For B$-$R versus MJD, $a = -0.059 \pm 0.002,\ b = 0.040 \pm 0.002,\ c = 0.106 \pm 0.006$ and $d = 0.809 \pm 0.004$, whereas for V$-$R versus MJD, $a =  -0.044 \pm 0.002,\ b =  0.040 \pm 0.002 ,\ c =  0.074 \pm 0.004$ and $d = 0.367 \pm 0.003 $. Although the long-term color versus JD plot shows slow variation fitted using a cubic polynomial, the segment-wise color versus JD variations are comparatively fast, although no definite trend is detected. Similarly to the color-magnitude plots, the color versus MJD trend becomes more evident for the more widely separated bands. 
 
\subsection{Z-transformed discrete correlation function}\label{zdcf}
\begin{figure*}
    \centering
    \includegraphics[width=0.92\linewidth]{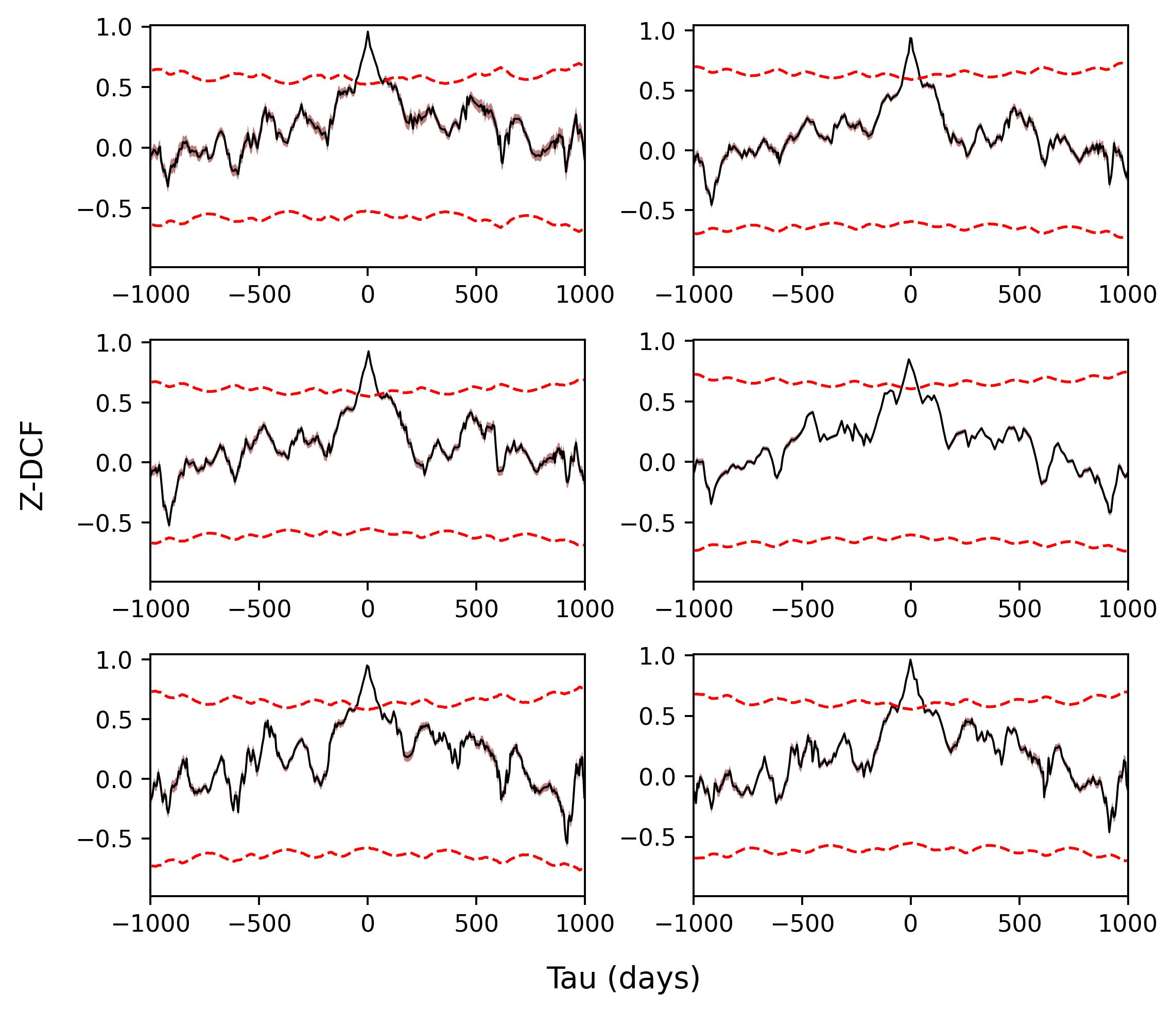}
    \caption{Cross correlation results between different optical bands. With all ZDCFs peaking at zero lag, one can infer essentially co-spatial emission, with the red dashed line representing 3.5$\sigma$ significance level.}
    \label{fig:dcf}
\end{figure*}

\noindent
To estimate the cross-correlation between these unevenly sampled time series, we employ the z-transformed discrete correlation function (ZDCF), originally introduced by \cite{1997ASSL..218..163A}. The ZDCF is an improvement over the discrete correlation function (DCF) developed by \cite{1988ApJ...333..646E}, addressing its statistical limitations by implementing two key modifications: equal-population binning and Fisher's z-transformation of the correlation coefficients. Unlike the traditional DCF, which bins time-lag pairs by fixed width, the ZDCF uses bins with an equal number of data pairs, improving statistical reliability across all lag ranges. 
Let $n$ be the number of $(x_i,y_i)$ pairs in a given time lag bin, then, for each bin, the correlation coefficient $r$ is computed as:

\begin{figure*}
    \centering
    \includegraphics[width=0.49\linewidth]{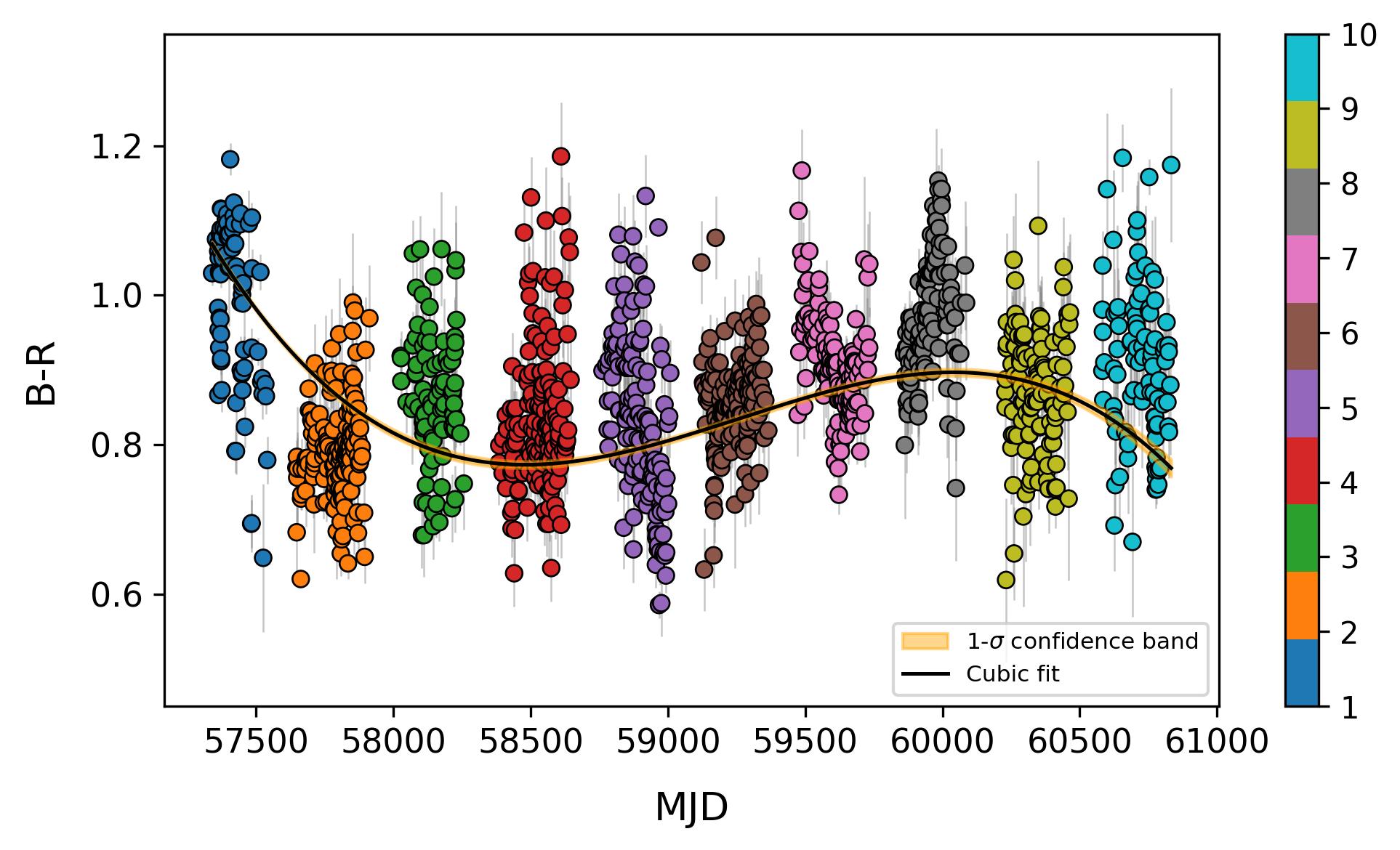}
    \includegraphics[width=0.49\linewidth]{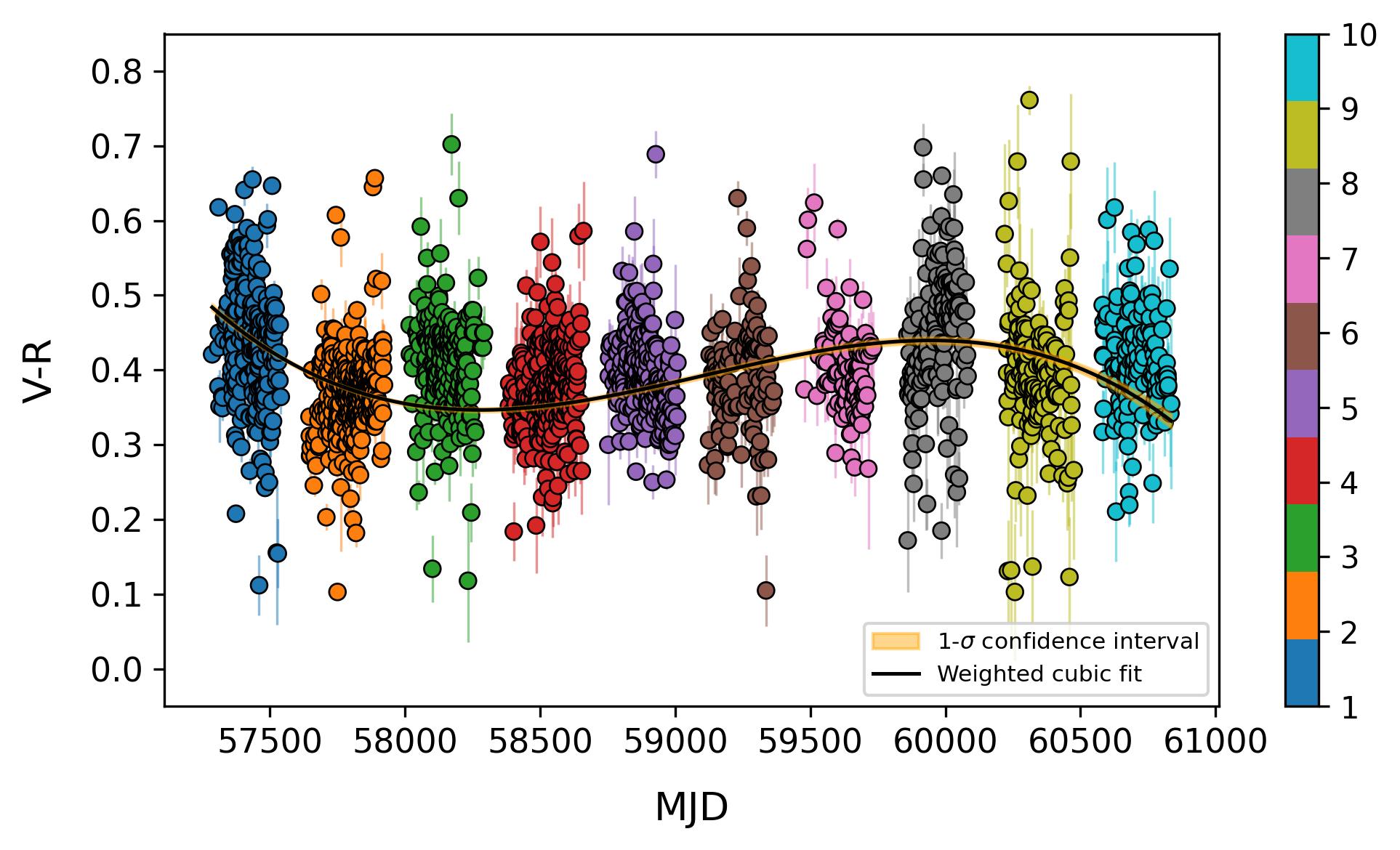}
    \caption{Color vs MJD plots for OJ 287. The plots show a slowly varying component, fitted with a cubic polynomial (in black).}
    \label{fig:color_vs_JD}
\end{figure*}

\begin{equation}
    r = \frac{\sum_i^n(x_i - \bar{x})(y_i - \bar{y})/(n-1)}{s_x s_y},
\end{equation}
which is then transformed using Fisher’s z-transformation:
\begin{equation}
    z=\frac{1}{2}\log \left( \frac{1+r}{1-r}\right), \hspace{0.2cm} \zeta = \frac{1}{2}\log \left( \frac{1+\rho}{1-\rho}\right), \hspace{0.2cm} r=\tanh z.
\end{equation}
where $\bar x, \bar y$ are bin averages, $s_x, s_y$ are their standard deviations, and $\rho$ is the unknown population correlation coefficient of the bin. Due to the presence of various emitting and absorbing regions surrounding an AGN, variations across different energy bands may not occur simultaneously, leading to observable time differences. To implement the ZDCF, we made use of the Python package {\tt pyzdcf\footnote{https://github.com/LSST-sersag/pyzdcf}} \citep{jankov_isidora_2022_7253034}. The results of the cross-correlations are shown in \autoref{fig:dcf}. As can be seen, the ZDCFs peak at zero lag, implying co-spatial optical emission. The significance of the detected peaks are shown by red dashed-lines in \autoref{fig:dcf} corresponding to a 3.5$\sigma$ level. The significance is calculated by simulating $5\times10^4$ light curves, following the procedure discussed in \cite{2013MNRAS.433..907E}.\\ 
\\
These results are consistent with not only the optical-optical correlations, but between different bands as well, although sometimes lags of few days have been detected for e.g., cross-correlation analyses have shown that optical and $\gamma$-ray light curves are often very closely aligned, with any lag mostly consistent with being zero or sometimes a few days \citep{2009ApJ...690.1018V, 2019MNRAS.486.1781R, 2021ApJ...923....7B}.  Such behavior tends to favor leptonic models where synchrotron emission and inverse-Compton scattering ($\gamma$-rays) come from the same lepton population, rather than more complex multi-zone or hadronic scenarios \citep{2009ApJ...697L..81B, 2014ApJ...783...83L, 2019MNRAS.490..124M}. Lags of the order of several to hundreds of days have also been observed, but that typically involves cross-correlations between, for example, lower frequency radio and optical light curves \citep{2019ApJ...887..185S}.

\section{Estimation of the Black Hole Mass in OJ 287 using Steward optical spectra}\label{sec:BH_mass}
\begin{figure*}
\centering
\includegraphics[width=0.48\linewidth]{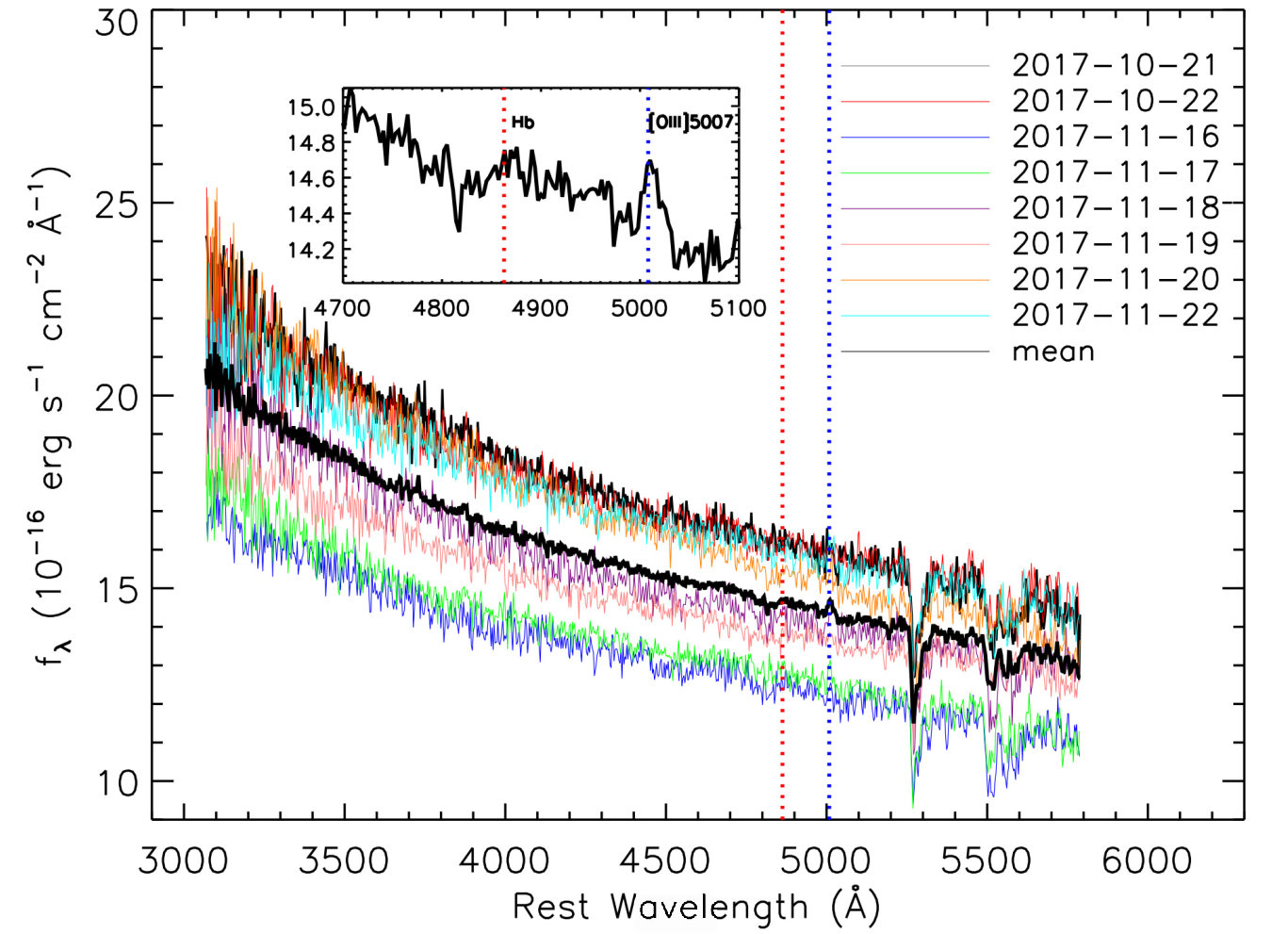}
\includegraphics[width=0.48\linewidth]{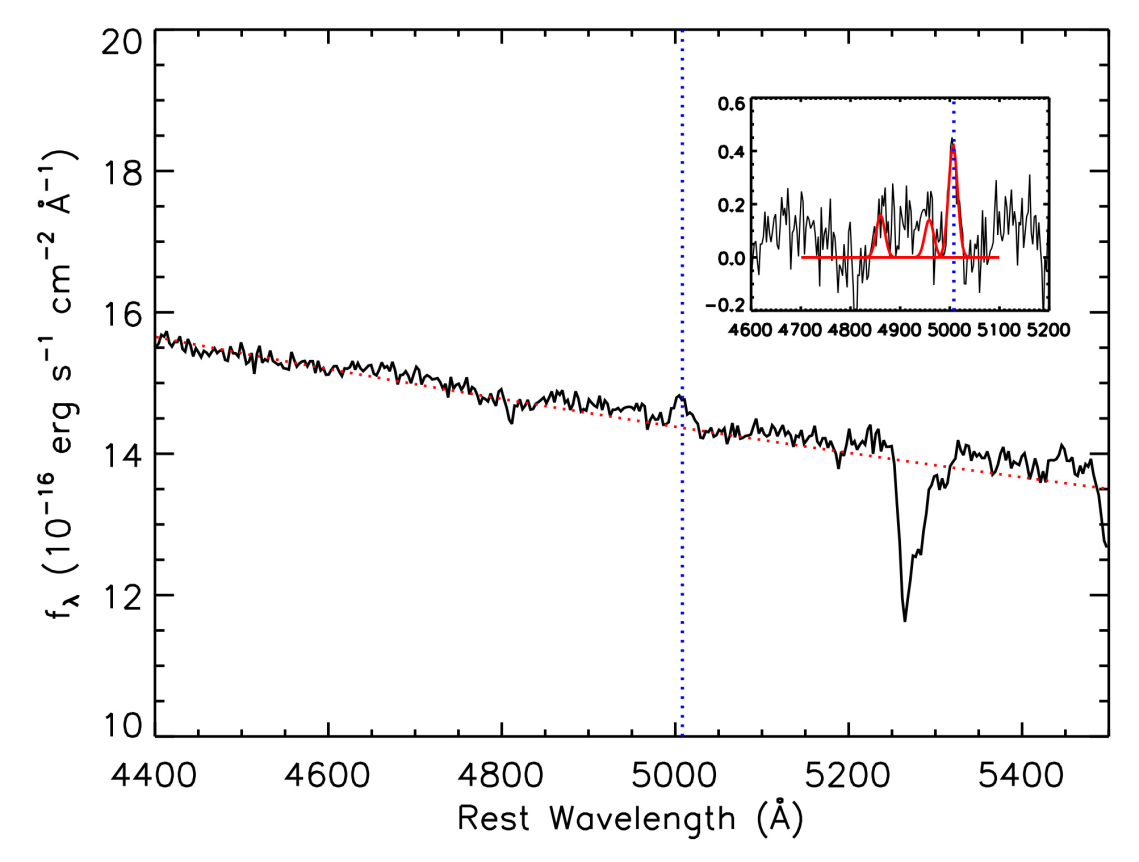} 
\vspace*{-0.2 cm}\caption{The eight spectra observed in the period of 2017.10.21-2017.11.22 and the flux average spectrum ({\it left}). All spectra were corrected for Galactic extinction and shifted to source rest-frame, with the position of emission lines of H$\beta$ and [O III] $\lambda 5007 \rm \AA$ indicated as the vertical red dotted and green dotted lines; the spectral fitting of OJ 287 ({\it right}), where the continuum is fitted with a single power law shown as the red dotted line, the black line is the original flux averaged spectrum and the inset shows the line fitting for [OIII] $\lambda 5007$\AA.}
\label{fig:spectra}
\end{figure*}

\noindent
We combined eight optical spectra of OJ 287 at its low state, in the observed period from 2017 October 21 to 2017 November 22, from Steward Observatory\footnote{http://james.as.arizona.edu/~psmith/Fermi} with the observed wavelength range $\lambda =$ (4000--7550)\AA. All eight spectra and the flux average spectrum are plotted in the left panel of \autoref{fig:spectra}.  
There are no significant emission lines in the individual spectra, but a weak [O~III] line appears at 5007 \AA \ in the average spectrum indicated by the blue dotted lines in both panels. As neither broad emission lines nor the host absorption lines are observed in the average spectrum,  the virial mass and M-$\sigma$ empirical relation estimations cannot be directly applied to obtain the black hole mass for OJ 287. \\
\\

However, we can employ the [O III] line width to estimate the black hole mass. The width of narrow line [O III] has frequently been used as a surrogate for $\sigma$ because the bulge gravity  dominates the global kinematics of the narrow-line region in AGN \citep[e.g.,][and references therein]{2007ApJ...667L..33K,2009MNRAS.398.1905W}. We analyzed the average spectrum following the procedure given in \citet{2020MNRAS.491...92L}, where the average spectrum was firstly corrected for Galactic extinction with the reddening map of \citet{1998ApJ...500..525S}, and then was shifted to the rest-frame wavelength by using the redshift of OJ 287. We modeled the continuum by applying a single power law ($f_{\lambda} \propto \lambda^{\alpha}$). The 
continuum-subtracted average spectrum was further fitted by using the Gaussian profile to fit the [OIII] $\lambda 5007 \rm \AA$ in the spectral range of 4700$-$5100 \AA. The final full width at half maximum (FWHM) of [O III] is 841 $\rm km~s^{-1}$, which was corrected for instrumental line broadening with the adopted source-frame instrumental resolution of FWHM$_{\rm instr} \simeq$ 918 $\rm km~s^{-1}$. In order to evaluate the corresponding error of FWHM, we generated 100 mock spectra by adding a random Gaussian noise to the original spectra using the flux density errors, and then calculated the standard deviation of the measurements from those mock spectra as the uncertainty. The flux density errors used here were the rms value of the average spectrum calculated over the source-frame spectral window of (5100$-$5150 \AA) after subtracting a second-order polynomial function. The spectral fitting is shown in the right panel of \autoref{fig:spectra}.\\

\noindent
Using the M-$\sigma$ relation of \citet{2013ARA&A..51..511K} where $\sigma = \rm FWHM_{\rm [OIII]}/2.35$, we estimated the BH mass to be log $(M_{\rm BH}/\rm M_{\odot}$)= 9.59. The systematic uncertainty of the M-$\sigma$ relation of 0.29 dex has been combined with the  [O III]-based FWHM (0.27 dex) to produce a total error for this black hole mass estimate, yielding log$~(M_{\rm BH}/\rm M_{\odot})= 9.59 \pm 0.40$. So the BH mass from this initial [O III] estimate could be up to $10 \times 10^{9}~\rm M_{\odot}$, which is comparable to that derived from the binary model for the massive one of $M_{\rm BH}$ = 18.35 $\pm ~0.05~\times$ 10$^{9}~\rm M_{\odot}$ \citep{2007ApJ...659.1074V}. The [O III] emission line is from the AGN narrow line emission region at the kpc scale, which is generally believed to be a cone structure rather than a disk; therefore, the projection effect may not be as important as when one considers the broad line region (BLR). The preferred lines for BH mass estimates, H$_{\alpha}$ and H$_{\beta}$ are from the BLR, which is located much closer to the BH than the narrow line emission region. There is another source of uncertainty which, while alluded to, is yet to be included. Since the line-emitting region is likely to be geometrically flat, and lie close to the plane of the sky in BL Lac objects such as OJ~287, the observed line width is narrower than it should be in comparison with an arbitrary viewing angle by a factor sin $i$ where $i$ is the inclination of the disk. Ideally, one would use an experimental the M-$\sigma$ relation for BL Lac objects, but since it is not available, the best we can do is to estimate the influence of the sin $i$ factor on the determined BH mass value. The correction factor could be up to a factor of ten for the NLR, so we can only say that the value of the BH mass mentioned above is a lower limit. The dynamically determined mass from the binary motion log ($M/M_{\odot}$) = 10.25 \citep{2007ApJ...659.1074V} is certainly compatible with this limit.\\
\\
We note that \cite{2023MNRAS.522L..84K} refer to \cite{2006ApJ...641..689V} who derived a correlation between the black hole mass $M_{BH}$ and the hydrogen line luminosity $L(H_{\beta})$ (in units of $10^{42} \ \rm{erg} \ \rm{s}^{-1}$) and the line width at half maximum $F(H_{\beta})$ (in units of 1000 km s$^{-1}$) in quasars:

\begin{equation}
M_{BH} = 10^{6.67} \ [L({H_{\beta})}]^{0.63} \ [F(H_{\beta})]^2 M_{\odot}.                 
\end{equation}

\noindent
They use the line parameters measured in 1984 December 23 by \cite{1985PASP...97.1158S} $L(H_{\alpha}) = 10^{42.8} \ \rm{erg} \ \rm{s}^{-1}$ and $F(H_{\alpha}) = 4200 \ \rm{km} \ \rm{s}^{-1}$ and obtain the BH mass of only, $M_{BH} \sim \rm{1.25} \times 10^{8} M_{\odot}$. They ignore the influence of the disk's inclination. The inclination of the accretion disk with respect to the plane of the sky $i$ is likely to be low in OJ~287, say $i \sim 5^{\circ}$, because the jet is pointing nearly in our direction. The Doppler broadening should be proportional to sin $i$, if the BLR is flattened. Comparing this with a typical value of (sin $i)^{-1}$ in a randomly oriented sample, this correction factor for OJ~287 would be around 10, increasing the estimated mass by a factor of $10^2$ over the correlation value. Therefore, after the inclination correction, the mass estimate of \cite{2023MNRAS.522L..84K} is consistent with what we find from the [O III] line, as well as the binary SMBH dynamical measurement. \\
\\
Now the binary SMBH model is based on the long-term quasi-periodic variability of the source, interpreted as the result of orbital motion and disk-impact events. Recent radio and multi-wavelength studies \citep[e.g.,][]{2024iSci...27j9427V,2025ApJ...992...60V} have identified multiple stable periodic components (of $\sim 12, \sim 6.5$, and $\sim 2.1$ yr), which are naturally explained within a binary framework and produce additional direct constraints on the system’s dynamical parameters. These analyses suggest a primary black hole mass of  $\sim 10^{10} M_{\odot}$ and a secondary component of $\sim 10^{9} M_{\odot}$, consistent with the dynamically determined mass from the long-term optical observations \citep[e.g.][]{2007ApJ...659.1074V}.\\
\\
Therefore, the lower mass derived from the $M–\sigma$ relation should be regarded as a conservative estimate, while the dynamically inferred mass from the binary model is most likely more representative of the true gravitational potential of the system. This is particularly relevant for OJ 287, where relativistic jet emission dominates and can significantly affect spectroscopic diagnostics. The comparison highlights the importance of combining spectroscopic and time-domain approaches when studying complex systems such as blazars hosting candidate binary SMBHs.

\section{Discussion}\label{sec:discussion}
\noindent
Here, we have investigated approximately ten years of densely sampled multi-band optical observations of OJ 287, spanning from 29th September 2015 to 26th April 2025. The dataset was compiled from an extensive network of telescopes worldwide, yielding the most densely sampled light curve of the source to date. It incorporates both previously published observations \citep{2017MNRAS.465.4423G, 2019AJ....157...95G, 2023ApJ...957L..11G} and newly obtained data provided by our collaborators. Additionally, publicly available data from the SMARTS and Steward observatories were included to further enhance the dataset's temporal coverage and reliability. Due to seasonal gaps, we divided the light curve into 10 distinct segments. All segments exhibit noticeable variability across all bands; however, the amplitude of variation differs significantly between them. Segment 1 (corresponding to 2015/6) shows the greatest variability, with brightness changes of approximately 2.7 mag, suggesting a phase of strong activity. In contrast, segment 4 (2018/9) shows the least variation, with a relatively modest change of $\sim$ 1 mag, illustrating a comparatively stable emission state. \\
\\
OJ 287 is a remarkable blazar that certainly appears to host a binary supermassive black hole system, where the secondary periodically impacts the accretion disk of the primary roughly every 12 years \citep{1996ApJ...460..207L, 1996A&A...315L..13S, 2008Natur.452..851V}. This periodic interaction produces characteristic optical outbursts, although none were captured in the present analysis, as our observational baseline is a few years shorter than the expected period.\\
\\
Blazar flux variability spans a wide range of timescales and can arise from both accretion-disk instabilities \citep{1997MNRAS.292..679L, 1991ApJ...376..214B, 2006MNRAS.367..801A} and jet-dynamics–driven processes. The former can lead to small-scale amplitude variations either due to instabilities in the accretion disk or by a change in accretion rate near the inner edge of the disk. Among the latter, the shock-in-jet model \citep{1985ApJ...298..114M} provides a natural explanation for many large flares in many wavebands, where shocks propagating along the relativistic jet accelerate particles and enhance synchrotron emission. The resulting emission can either predominantly originate from a single zone \citep{1998MNRAS.301..451G, 1998ApJ...509..608T}, or seems to require contributions from multiple zones \citep{2011MNRAS.416.2368C, 2014ApJ...780...87M}. Collisions between stationary and traveling shocks within the jet give rise to multiple radiation zones and are used to explain the short-term multi-wavelength variability; meanwhile, the single zone models are helpful in explaining the long-term variability since the finite size of the emission region and the associated light-crossing time, along with particle injection timescales, naturally predict a larger minimum timescale for flaring episodes \citep[e.g.][]{2020AN....341..713S}. Geometrical effects can also play a key role in observed variability by affecting the bulk Lorentz factor, which in turn can increase or decrease the observed flux \citep{2017Natur.552..374R, 2019ApJ...887..185S}.\\
\\
Blazars typically exhibit two main types of color behavior: redder-when-brighter (RWB) and bluer-when-brighter (BWB) trends. As a blazar, and specifically a BL Lac object in which quasi-thermal accretion disk emission is minimal, the variability of OJ 287 is primarily driven by the jet emission. In the case of BL Lac objects, most studies point toward a dominant BWB trend \citep{2006A&A...450...39G, 2013MNRAS.432.1189Z}. In this work, we analyzed the long-term color variability of OJ 287 and found a clear BWB trend, not only in the full light curve, but also consistently across all ten individual segments. The BWB trend seen in BL Lac objects can be attributed to electrons accelerated by the shock front in the jet losing energy more quickly, leading to greater variability in the higher-energy bands \citep{2002PASA...19..138M}. Alternatively, the rise in luminosity could be explained by the injection of new electrons with a harder energy distribution compared to the previously cooled population.\\
\\
Since the confirmation of the large precession in the binary orbit and the detection of loss of the gravitational wave energy from the system \citep{2008Natur.452..851V}, the interest in multi-wavelength observations has boomed. In particular, the predicted 2015 flare was already much better covered observationally than the previous flares that mark the binary orbit \citep{2016ApJ...819L..37V}. This could be considered the start of the high MW observational efforts. As far as we know, the spectral and temporal behavior has not been any different from earlier flaring periods, but the coverage has improved \citep{2018MNRAS.479.1672K,2018MNRAS.473.1145K}. The dynamic evolution of the emission from OJ~287, backed by a certain predictability, rare in a transient dynamical system, has encouraged a great deal of MW monitoring. Investigations of these accumulated rich data sets have led to the discovery of many new features including: a thermal spectral component in the NIR-optical band \citep{2018MNRAS.473.1145K}; an additional HBL-like jet emission component \citep{2017ICRC...35..650B,2018MNRAS.479.1672K, 2021ApJ...921...18K, 2022MNRAS.509.2696S}, with an unusually soft X-ray spectrum \citep{2017IAUS..324..168K,2020MNRAS.498L..35K,2018MNRAS.473.1145K, 2021ApJ...921...18K,2018MNRAS.480..407K,2021ApJ...920...12H}; a Seyfert-like UV and soft-X-ray excess \citep{2020ApJ...890...47P}; an iron absorption feature \citep{2020MNRAS.498L..35K}; systematic trends in optical polarization \citep{2018ApJ...862....1C,2019AJ....157...95G,2023ApJ...957L..11G}; and secondary jets \citep{2024sf2a.conf..377B,2025ApJ...992..110V}, most of these for the first-time. Interestingly and intriguingly, all of these newer, peculiar features are detected when the two SMBHs are closely interacting as per the dynamical evolution of the disk-impact model. These, in turn, has further expanded the scope of observations and now OJ 287 is one of the targets of the Event Horizon Telescope \citep[EHT;][]{2024A&A...683A.248C}.\\
\\
These fascinating findings in turn have greatly enlarged the physics of this source that can be probed, apart from binary aspects and jet/accretion. These include: strong gravity physics involving tidal disruptions \citep{2021ApJ...920...12H}; BH properties \citep[e.g.][and references therein]{2008Natur.452..851V,2023MNRAS.525.1153V,2023ApJ...957L..11G}; gravitational waves \citep[e.g.][and rererences therein]{2008Natur.452..851V,2018MNRAS.481.2249C,2023ApJ...951L...8A}; accretion and aspects of relativistic jets \citep[e.g.][and references therein]{2020MNRAS.498L..35K,2022ApJS..260...39G,2023ApJ...957L..11G,2025Univ...11...84K, 2022JApA...43...79K, 2025ApJ...993L..22R}, aspects of cosmology \citep[e.g.][]{2025ApJ...984L..66D,2024ApJ...962L..40C,2024Galax..12...34Y}; and finally, as a multi-messenger source \citep[e.g.][]{2020MNRAS.498.5424R,2021Galax..10....1V}. 

\section{Conclusions}\label{sec:summary}
\noindent
This study presents densely sampled investigations of the long-term optical variability of the blazar OJ 287. The analysis is based on an extensive collection of optical photometric observations acquired from numerous ground-based telescopes worldwide, covering its emission as continuously as possible from the ground between 2015 and 2025. A concise summary of our main results is provided below.
\noindent
\begin{enumerate}

\item Due to the seasonal gaps, the light was divided into a total of 10 segments, with the object going through a maximum and minimum magnitude change of $\sim$ 2.87 and 0.86 in V-band (segment 1) and I-band (segment 4), respectively.  The object was found to be variable throughout the observing campaign, but faded steadily after the peak of the 2015 disk impact flare, reaching V=16.86 in April 2025.

\item The color–magnitude relationship for OJ~287 was analyzed both over the entire light curve and across individual segments. A clear “bluer-when-brighter” trend was observed, not only in the full light curve, but also consistently within each segment. Among the B$-$R color indices, the strongest and weakest correlations were found in segments 5 (2019/2020) and 9 (2023/2024), with Spearman correlation coefficients, $\rho$, of 0.717 and 0.258, respectively. For the full light curve, Spearman $\rho$ was 0.495.

\item The discrete correlation function analysis was performed between all bands using ZDCF. All inter-band cross-correlations were found to peak at zero lag, implying that the emission is co-spatial in the BVRI bands.

\item We combined eight optical spectra of the source to estimate its black hole mass by using the width of the [O III] line as a surrogate for $\sigma$ in the M--$\sigma$ relation. The continuum was fitted using a power law, while the continuum-subtracted average spectrum was fitted using a Gaussian function, giving a FWHM$_{instru} \simeq$ 918 $\rm km~s^{-1}$ which translates to a black hole mass of log $M_{\rm BH} = \rm{9.59} \pm \rm{0.40} \rm{M}_{\odot}$. This should be considered as the lower limit for the mass of the primary black hole; it is consistent with the dynamically determined mass of log ($M/M_{\odot}$) = 10.25.

\end{enumerate}

\section*{ACKNOWLEDGMENTS}

\noindent
We are sad to note that V.~M.\ Larionov, R.\ Pickard, and E.\ Semkov passed away during this extended observation campaign. They were dedicated collaborators, and their contributions to our team will be greatly missed. We thankfully acknowledge the contribution by V.~M.\ Larionov, who is not included as a co-author of the paper at the suggestion of his family. \\
\\
Data from the Steward Observatory spectropolarimetric monitoring project were used. This program is supported by Fermi Guest Investigator grants NNX08AW56G, NNX09AU10G, NNX12AO93G, and NNX15AU81G. This paper has made use of up-to-date SMARTS optical/near-infrared light curves that are available\footnote{http://www.astro.yale.edu/smarts/glast/home.php}. \\
\\
We thank the anonymous reviewer for constructive comments. This work was supported by the Tianshan Talent Training Program with Grant No.~2023TSYCCX0099. ACG is partially supported by the CAS ``President's International Fellowship Initiative (PIFI)" with Grant No.~2026PVA0040. The Abastumani team acknowledges financial support by the Shota Rustaveli NSF of Georgia under contract FR-24-515. PK acknowledges support from the Department of Science and Technology (DST), government of India, through the DST-INSPIRE Faculty grant (DST/INSPIRE/04/2020/002586). The research at Boston University was supported in part by several NASA Fermi Guest Investigator grants, the latest is 80NSSC26K0222. This study was based in part on observations conducted using the 1.8m Perkins Telescope Observatory (PTO) in Arizona, which is owned and operated by Boston University.  KM acknowledges support from JSPS KAKENHI grant number 19K03930. The work of SMH and XC is supported by the National Natural Science Foundation of China under grant No. 12373015. GD, OV, MS and MDJ acknowledge support by the Astronomical Station Vidojevica and the Ministry of Science, Technological Development and Innovation of the Republic of Serbia (MSTDIRS) through contract no. 451-03-33/2026-03/200002 made with Astronomical Observatory (Belgrade), by the EC through project BELISSIMA (call FP7-REGPOT-2010-5, No. 256772), the observing and financial grant support from the Institute of Astronomy and Rozhen NAO BAS through the bilateral SANU-BAN joint research projects ``Astrometric and astrophysical research of variable astronomical objects", and ``Optical flux stability of 47 Active Galactic Nuclei important for the link between the Gaia CRF and ICRF systems", and support by the SANU project F-187. Also, this research was supported by the Science Fund of the Republic of Serbia, grant no. 6775, Urban Observatory of Belgrade - UrbObsBel, and grant no. 7337 Modeling Binary Systems That End in Stellar Mergers and Give Rise to Gravitational Waves - MOBY. This research was partially supported by the Bulgarian National Science Fund of the Ministry of Education and Science under grant KP-06-H88/4 (2024). ML is supported by the grants from the Rubin-Chile Fund (DIA3324). JHF's work partially supported by the National Natural Science Foundation of China (NSFC grant 12433004) and grant 2025YFA1614102. HG acknowledges financial support from the Department of Science \& Technology, India through INSPIRE faculty award IFA17-PH197 at ARIES, Nainital. MFG is supported by the National Science Foundation of China (grant 12473019), the Shanghai Pilot Program for Basic Research-Chinese Academy of Science, Shanghai Branch (JCYJ-SHFY-2021-013), the National SKA Program of China (Grant No. 2022SKA0120102), the National Science and Technology Major Project (2024ZD1100601), and the China Manned Space Project with No.\ CMS-CSST-2025-A07. BV is funded by the Swedish Research Council (Vetenskapsr\aa det, grant no. 2017-06372). ZZ is thankful for support from the National Key R\&D Programme of China (under grant no. 2018FA0404602) and the Chinese Academy of Sciences (CAS) for the Talented Program. WWZ is supported by the Shanghai Natural Science Foundation Youth Project 25ZR1402546, and Strategic Priority Research Program of the Chinese Academy of Sciences (CAS) Grant No.\ XDB0800302.

\section*{Data availability}
\noindent
Data may be provided on a reasonable request to ACG or KD.

\bibliography{oj287_optical_refs}{}
\bibliographystyle{aasjournalv7}

\end{document}